\documentclass[twocolumn,nodate,showpacs,preprintnumbers,nofootinbib,amsmath,amssymb,aps,prd]{revtex4}

\usepackage{graphicx}
\usepackage{amsmath,amssymb}
\usepackage[usenames]{color}

\newcommand{\half}{\frac{1}{2}}
\newcommand{\halfi}{\frac{i}{2}}
\newcommand{\third}{\frac{1}{3}}
\newcommand{\sgn}{\operatorname{sgn}}

\begin{document}

\title{Corrections to the Friedmann Equations from LQG \\ for a Universe with a Free Scalar Field }
\author{Victor Taveras}
\email{victor@gravity.psu.edu}
\affiliation{Institute for Gravitation and the Cosmos \\
			 The Pennsylvania State University \\
			 University Park, PA 16802-6300, USA}

\begin{abstract}
In loop quantum cosmology the quantum dynamics is well understood.  We approximate
the full quantum dynamics in the infinite dimensional Hilbert space by projecting it 
on a finite dimensional submanifold thereof, spanned by suitably chosen semiclassical 
states.  This submanifold is isomorphic with the classical phase space and the 
projected dynamical flow provides effective equations incorporating the leading 
quantum corrections to the classical equations of motion.  
Numerical work has been done using quantum states which are semiclassical at late times.
These states follow the classical trajectory until the density is on the order of 1\% of the
Planck density then deviate strongly from the classical trajectory.  The effective
equations we obtain reproduce this behavior to surprising accuracy.
\end{abstract}
\preprint{IGC-08/7-2}
\maketitle

\section{Introduction}
Loop Quantum Gravity (see [1-3] 
for reviews) is a nonperturbative 
approach to the problem of quantizing gravity.  An open problem is that of the semiclassical 
limit, i.e. are there solutions to LQG which closely approximate solutions to the classical
Einstein's equations?  Although this remains an open problem, concrete results can be achieved in the context of
Loop Quantum Cosmology (for recent review see [4,5]) in the spatially flat case.  
Loop Quantum Cosmology is a symmetry reduction of LQG, i.e. a quantization of a symmetry reduced sector of general
relativity.  There, it is the case that one can indeed find semiclassical
solutions in the quantum theory that closely approximate solutions to the classical Einstein
equations, namely, the Friedmann equations at late times.  Our goal is to to not just find 
semiclassical solutions but we would like to be able to go further and find effective equations 
incorporating the leading quantum corrections.

It is a common misconception that canonically quantizing general relativity would just reproduce Einstein's
equations without any modifications.  The classical equations are in fact modified by
quantum corrections.  Using the geometric quantum mechanics framework \cite{ts} this has been previously 
shown for a dust-filled, spatially flat Friedmann universe \cite{jw} in the so-called $\mu_0$ 
framework (see reference 3 in \cite{aps}).  In this paper we work in the so-called $\bar{\mu}$ framework and show that 
this is also the case for a spatially flat Friedmann universe with a free scalar field.  

We can find effective equations arising from the geometric quantum mechanics framework \cite{ts}
(`effective equations' for short). These semiclassical states follow the classical trajectory 
until the scalar field density is on the order of 1\% of the Planck
density where deviations from the classical trajectory start to occur. Then there are
major deviations from the classical theory. However, states which are semiclassical at late
times continue to remain sharply peaked \emph{but} on the trajectory given by the `effective equations'. Thus,
we can approximate the full quantum dynamics in the (infinite dimensional) Hilbert space by a system 
of `effective equations', incorporating the leading quantum corrections, on a finite dimensional submanifold 
thereof isomorphic to the classical phase space.  Furthermore, we know from comparison with the numerical work 
that the leading corrections are sufficient to capture essential features of the full quantum dynamics \cite{aps}.

We begin by briefly summarizing the classical theory in \ref{sec:CT}.  We then provide an
overview of the framework for effective theories that we use to obtain the `effective equations'
in \ref{sec:FfET}.  Then in the quantum theory, \ref{sec:QT} we choose a family of
candidate semiclassical states, verify that they are indeed sharply peaked, and compute the
effective Hamiltonian constraint.  We obtain the `effective equations' in \ref{sec:EE} and
show that they are self-consistent to within our order of approximation.  Finally, we discuss how much 
this approximation can be trusted near the bounce point in the Appendix and conclude.

Additionally, in this paper we use the choice of convention: $l_p^2 = G\hbar$, $c_{light}=1$, $\kappa = 8\pi G$, and
FRW parameter $k=0$ since we work in the spatially flat case.  
\section{Classical Theory}\label{sec:CT}
In the classical theory, the phase space $\Gamma$ is consists of pairs ($A_a^i$,$E^a_i$)
where $A_a^i$ is an $su(n)$ connection and the $E^a_i$ its canonically conjugate field. 
Due to isotropy and homogeneity we have a fiducial background triad $\;{}^oe^a_i$ and 
connection $\;{}^o\omega_a^i$.  Additionally, we may fix a fiducial cell with volume
$^oV$ and restrict our calculations to that cell.  There, we can write $A_a^i$ and $E^a_i$
as \cite{amj} 
\begin{eqnarray}
	\label{eq:A}
	A_a^i &=& c\;{}^oV^{-\frac{1}{3}} \;{}^o\omega_a^i \\ 
	\label{eq:E}
	E^a_i &=& p\sqrt{^oq}\;{}^oV^{-\frac{2}{3}} \;{}^oe^a_i
\end{eqnarray}
where $^oq$ is the determinant of the fiducial metric. Due to homogeneity and isotropy 
all the nontrivial information in $A_a^i$ and $E^a_i$ is contained in the variables $c$ and 
$p$.  The Hamiltonian constraint in these variables for a Friedmann universe with a free scalar 
field $\phi$, in the regime $p>0$ is given by,
\begin{equation}
	\label{eq:1a}
	C=-\frac{3}{\kappa \gamma^2}c^2p^\frac{1}{2}+\frac{1}{2}\frac{p_\phi^2}{p^\frac{3}{2}}=0
\end{equation}
where $p_\phi$ is the conjugate momentum of the massless scalar field $\phi$.

For convenience we switch to variables given by $\beta=c/\sqrt{p}$
and $V=p^\frac{3}{2}$ in which the Hamiltonian is given by
\begin{equation}
	\label{eq:1}
	C=-\frac{3}{\kappa\gamma^2}\beta^2V+\half\frac{p_\phi^2}{V}=0
\end{equation}
These new canonically conjugate variables are related to the old
geometrodynamics variables via
\begin{eqnarray}
	\label{eq:2}
	\beta = \gamma \frac{\dot{a}}{a} , \\
	\label{eq:3}
	V = a^3 ,
\end{eqnarray}
With the inclusion of a scalar field, the symplectic structure now has an extra part,
\begin{equation}
	\label{eq:4}
	\Omega=\frac{2}{\kappa\gamma}d\beta\wedge dV+d\phi \wedge dp_\phi ,
\end{equation}
where the scalar field is $\phi$ and it's conjugate momentum is $p_\phi$.  The Poisson 
bracket on the phase space is given by,
\begin{equation}
	\label{eq:5}
	\{f,g\}=\frac{\kappa\gamma}{2}\left(\frac{\partial f}{\partial \beta}\frac{\partial g}{\partial V}-\frac{\partial g}{\partial \beta}\frac{\partial f}{\partial V}\right)+\frac{\partial f}{\partial \phi}\frac{\partial g}{p_\phi}-\frac{\partial g}{\partial \phi}\frac{\partial f}{\partial p_\phi} .
\end{equation}
where the phase space $\Gamma$ consists of all possible points $\{\beta,V,\phi,p_\phi\}$.  The
allowed pairs of points in $\Gamma$ are those satisfying the Hamiltonian constraint
\begin{equation}
	\label{eq:6}
	C(\beta,V,\phi, p_\phi ) = 0.
\end{equation}

Recall, that for a function that depends on the canonical variables, its time 
dependence is given by its Poisson bracket with the Hamiltonian.  Thus the 
classical equations of motion are given by:
\begin{eqnarray}
	\label{eq:7}
	\dot{\beta}=\{\beta,C\} &=& -\frac{3}{2}\frac{\beta^2}{\gamma}-\frac{\kappa\gamma}{4}\frac{p_\phi^2}{V^2}, \\
	\label{eq:8}
	\dot{V}=\{p,C\} &=& 3\frac{\beta}{\gamma}V, \\
	\label{eq:9}
	\dot{\phi}=\{\phi,C\} &=& \frac{p_\phi}{V}, \\
	\label{eq:10}
	\dot{p_\phi}=\{p_\phi,C\} &=& 0.
\end{eqnarray}
We can verify that (\ref{eq:7}) and (\ref{eq:8}) are equivalent to the Friedmann equations for a free scalar 
field when written in terms of ordinary ADM variables.  Looking first at (\ref{eq:8}) 
we see that it gives that,
\begin{equation}
	\label{eq:11}
	\beta = \gamma\frac{\dot{a}}{a}.
\end{equation}
Putting in this expression in (\ref{eq:6}) we obtain,
\begin{equation}
	\label{eq:12}
	-\frac{3}{\kappa}\dot{a}^2a+\frac{1}{2}\frac{p_\phi^2}{a^3}=0,
\end{equation}
which can be rewritten as the Friedmann's equations,
\begin{equation}
	\label{eq:13}
	3\frac{\dot{a}^2}{a^2}=\kappa\left(\frac{1}{2}\frac{p_\phi^2}{a^3}\right)\frac{1}{a^3}.
\end{equation}
or in terms of the Hubble parameter and the scalar field density,
\begin{equation}
	\label{eq:13b}
	H^2=\frac{\kappa}{3}\rho
\end{equation}
Now putting in the equation for $\beta$ into (\ref{eq:6}) we can obtain the Raychaudhuri equation,
\begin{equation}
	\label{eq:14}
	3\frac{\ddot{a}}{a}=-2\kappa\rho
\end{equation}
Now \eqref{eq:9} and \eqref{eq:10} give us, respectively,
\begin{equation}
	\label{eq:15}
	\dot{\phi}=\frac{p_\phi}{p^\frac{3}{2}},
\end{equation}
and
\begin{equation}
	\label{eq:16}
	p_\phi=\text{const},
\end{equation}
since $C$ does not depend on $\phi$ for a free scalar field.  These are just the equations of motion for a free scalar field.
It is to the Hamiltonian constraint and these 4 equations, i.e. \eqref{eq:7}, \eqref{eq:8},
\eqref{eq:9}, \eqref{eq:10}, that we wish to find the corrections due to quantum gravity effects.
\section{Framework for Effective Theories}\label{sec:FfET}
Fortunately, there is the so-called geometric quantum mechanics framework which gives us a framework
in which we can obtain these effective equations. In this section we briefly review the framework for 
effective theories.  A more through review may be found in \cite{jw} and \cite{ts}.  This framework is especially suitable because it
provides a direct route to the effective equations from the Hamiltonian constraint without having
to deparameterize the theory.  That is, in background independent theories there is no canonical notion of time so one 
usually deparameterizes the theory by choosing one of the fields as an internal clock and then proceeds
by considering the dynamics of the other fields with respect to the field serving as a clock.  In particular,
this model is deparamaterized and analyzed in \cite{aps} using the scalar field $\phi$ as an internal clock.

We take a brief detour and present the main idea behind the procedure for the simpler example of a particle moving in 
a potential along the real line.  The proof that quantum mechanics has the correct semiclassical 
limit is usually based on an appeal to Ehrenfest's theorem, which is usually expressed in the form,
for a particle in a potential,
\begin{equation}
	\label{eq:17}
	m \frac{\partial^2 \langle x \rangle}{\partial t^2} = - \left\langle \frac{\partial V(x)}{\partial x}\right\rangle .
\end{equation}
However, this holds for all states, not just semiclassical ones.  To recover classical
equations for $\langle x\rangle$ what we would like 
to do is be able to pull the derivative outside of the expectation value,
\begin{equation}
	\label{eq:18}
	m \frac{\partial^2 \langle x \rangle}{\partial t^2} \simeq -\frac{\partial V(\langle x \rangle)}{\partial \langle x \rangle} .
\end{equation}
Where in pulling the derivative outside, we have replaced $\frac{\partial}{\partial x}$ 
with $\frac{\partial}{\partial \langle x \rangle}$, since after taking the expectation 
value the right hand side is now not a function of $x$ but of $\langle x \rangle$.  
Additionally, we would like to know if we are able to express the corrections to the 
right hand side of \eqref{eq:18} in terms of a corrected potential,
\begin{equation}
	\label{eq:19}
	m \frac{\partial^2 \langle x \rangle}{\partial t^2} = -\frac{\partial}{\partial \langle x \rangle}\left[V(\langle x \rangle) + \delta V(\langle x \rangle) \right] + \ldots .
\end{equation}
That is, we are looking for an equation of motion satisfied by the expectation values
which includes quantum corrections to the classical equations of motion.
A priori, it is not certain if we can do this.  However, the geometrical formulation of 
quantum mechanics provides a framework in which we can answer this question. Therefore,
we quickly review this framework.  A more thorough review can be found in \cite{ts}.

In the quantum theory quantum states are represented by elements of a Hilbert space 
$\mathcal{H}$. More precisely, the quantum phase space consists of rays in $\mathcal{H}$. 
The space $\mathcal{H}$ can be made into a symplectic space via its 
inner product, which can be decomposed into its real and imaginary parts,
\begin{equation}
	\label{eq:21}
	\langle \phi | \psi \rangle = \frac{1}{2\hbar}G(\phi, \psi) + \frac{i}{2\hbar}\Omega(\phi, \psi) ,
\end{equation}
where $G$ gives a Riemannian metric and $\Omega$ a symplectic form on $\mathcal{H}$. Given
a function $f$ on the phase space, one can construct the Hamiltonian vector field associated
with $f$ by using the symplectic form,
\begin{equation}
	\label{eq:hvf}
	X_{(f)}^a=\Omega^{ab}\nabla_bf
\end{equation}
It turns out that the flow generated on the phase space by the Hamiltonian vector field $X_{(f)}^a$
corresponds to Schrodinger evolution in the quantum theory generated by $\hat{f}$ \cite{ts}.

Taking expectation values of the operators corresponding to the canonical variables
provides a natural projection, $\pi$, from states in the Hilbert space to points in the 
classical phase space.  Thus, the Hilbert space, $\mathcal{H}$ is naturally viewed 
as a fiber bundle over the classical phase space $\Gamma$,
\begin{equation}
	\label{eq:20}
	\pi : \mathcal{H} \rightarrow \Gamma .
\end{equation}
Because of the fiber bundle structure, the classical phase space can be viewed 
as a cross section of the Hilbert space.  

We can use this symplectic form to define a notion of horizontal vectors in $\mathcal{H}$.
Namely, vertical vectors are vectors whose components are in the directions in which
the expectation values don't change and horizontal vectors are vectors orthogonal to the 
vertical vectors.  It turns out that one has to use $\Omega$ rather than $G$ to define 
orthogonality.  That is, two vectors, $\phi$ and $\psi$, are orthogonal if
\begin{equation}
	\label{eq:22}
	\Omega(\phi, \psi) = 0 .
\end{equation}
To summarize, $\Omega$ gives us a notion of horizontal vectors which we use to construct
horizontal sections of the Hilbert space.  

Because of the fiber bundle structure, \textit{any} horizontal section can be identified with 
the classical phase space.  At the kinematical level there is no natural section for 
us to choose.  However, later when  we consider dynamics in section \ref{sec:EE} we can 
look for a natural section that is approximately preserved by the flow of the Hamiltonian 
constraint in a precise sense.

A priori, we have no way of knowing whether such a section exists.
However, \textit{if} we can find such a section then the quantum dynamics on such a 
section can be expressed in terms of an effective Hamiltonian which is simply the 
expectation value of the quantum Hamiltonian operator \cite{ts}.  The expectation value yields 
the classical term as the leading term and gets corrections due to quantum effects in 
the subleading terms.  This is the key idea behind our calculation.

This can be done exactly for, e.g., the harmonic oscillator and approximately for several
other physically interesting systems, including a dust-filled Friedmann universe \cite{jw}.  The 
main result of this paper is proving that this can also be done approximately for a Friedmann 
universe with a free scalar field and thus obtaining effective equations for this model. In
the remainder of this work we will call these 'effective equations' because they are not effective 
equations in the traditional sense.  These are on a different footing
from traditional effective equations since there is no precise definition of approximately horizontal 
sections.  Nonetheless, we show that we can obtain these `effective equations' in a well-controlled
approximation
\section{Quantum Theory}\label{sec:QT}
Recall that in the full theory the elementary variables are the holonomies
of the connection and the electric fluxes.  Recall that in loop quantum cosmology
there exists no operator $\hat{c}$ \cite{amj} corresponding to the connection, and thus 
$\hat{\beta}$ also by extension.  However, the holonomy operator,
$\exp(\halfi\bar{\mu}c)=\exp(\halfi\sqrt{\Delta}\beta)$ does exist. Therefore in 
the quantum theory we work with the algebra generated by $\exp(\halfi\sqrt{\Delta}\beta)$
and $\hat{V}$
\subsection{Coherent State}\label{CS}

In the previous section, we saw that the taking of expectation values provides a 
natural projection from the Hilbert space to the classical phase space.  These
expectation values can be taken in any state.  However, for our investigation of
the semiclassical limit it is natural to choose a semiclassical state.  Simplest
candidates are the Gaussian coherent states. They are natural because we can choose them to be sharply peaked at classical values of 
the canonical variables, i.e. with small spread in both canonically conjugate variables.  
Let $v$ and $v'$ denote the parameters proportional to the 
eigenvalues of the volume operator, related to the volume eigenvalues $V$ and $V'$ 
via $V=(\frac{8\pi\gamma}{6})^\frac{3}{2}\frac{l_p^3}{K}v$  and 
$V'=(\frac{8\pi\gamma}{6})^\frac{3}{2}\frac{l_p^3}{K}v'$ where $K=\frac{2\sqrt{2}}{3\sqrt{3\sqrt{3}}}$.  
Then a Gaussian coherent state $(\psi|$, with Gaussian spreads $\epsilon$ and $\epsilon_\phi$ 
in the gravitational sector and scalar field sector of the state, peaked at some classical values 
$\beta',V',\phi',p_\phi'$ is given by,
\begin{eqnarray}
	\label{eq:23}
	(\psi_{\beta',V';\phi',p_{\phi}'}|&=&\int dp_\phi\sum_v e^{-\frac{1}{2}\epsilon^2(v-v')^2}e^{\halfi\sqrt{\Delta}\beta'(v-v')}\nonumber \\
	&&\times e^{-\frac{1}{2}\epsilon_\phi^2(p_\phi-p_\phi')^2}e^{-i\phi'(p_\phi-p_\phi')}(v;p_\phi| \nonumber \\
	&& =: \int dp_\phi \sum_v \overline{\psi_n}(p_\phi) (v;p_\phi| . 
\end{eqnarray}
As usual, $\psi$ is defined on a lattice and the summation index $v$ runs over the integers.
It is important to note that the spread $\epsilon$ is not constant but is a function of the 
phase space point.  We will provide the reason and more details on the functional dependence in the 
next section.  Recall that solutions to the constraints do not lie in the kinematical Hilbert space, 
but rather in its algebraic dual \cite{almmt}. Physical states, such as the semiclassical state given in 
\eqref{eq:23} should thus lie in the dual space and thus we write it as a so-called `bra' state 
$(\psi|$.  At first it would seem that working in the dual space would be unmanageable but it 
does not pose computational difficulties. However, because one fortunately has the ``shadow state
framework" to carry out calculations \cite{afw}-\cite{al2}.

We will use $(\psi|$ as our semiclassical state and calculate all of our expectation
values in this state.    The interpretation of the basis ket is that the universe 
in the state $|v;p_\phi\rangle$ has physical volume $v$ in Planckian units and scalar
field momentum $p_\phi$. Our choice for $(\psi|$ as a Gaussian coherent state is not the 
most general but, rather, is the `simplest' choice one can make to obtain 
these `effective equations' in a late-time, large-volume approximation.  For
more general considerations it is possible to characterize the state in terms of all its
`moments' as in \cite{mbeff}.  The higher moments acquire their own equations of motion
which must be solved simultaneously with the equations of motion for the canonical variables.

\subsection{Operators in the Quantum Theory}\label{sec:OinQT}
In order to take expectation values, we need to define the operators in the quantum
theory corresponding to the canonical variables.   We now construct an approximate 
operator for $\beta$ in terms of exponentiated $\beta$ variables in order
to verify that the state $\Psi$ is indeed sharply peaked at classical value of $\beta'$.
Our point of departure is the classical expression
\begin{equation}
	\label{eq:24}
	\beta \simeq \frac{1}{i\sqrt{\Delta}}\left(e^{\halfi\sqrt{\Delta}\beta}-e^{-\halfi\sqrt{\Delta}\beta}\right)
\end{equation}
which is exact in the limit  $\sqrt{\Delta}\beta \rightarrow 0$. However, our experience
in the full theory tells us that in the quantum theory we should not be taking this limit to 0 but to the area 
gap $\Delta=4\sqrt{3}\pi\gamma l_p^2$. So in our quantum theory we take 
\begin{equation}
	\label{eq:25}
	\hat{\beta}_\Delta = \frac{1}{i\sqrt{\Delta}}\left(\widehat{e^{\halfi\sqrt{\Delta}\beta}} - \widehat{e^{-\halfi\sqrt{\Delta}\beta}}\right).
\end{equation}
Thus, our operator $\hat{\beta}_\Delta$ agrees approximately with the classical $\beta$ in
the regime $\sqrt{\Delta}\beta \ll 1$.  The choice for $\hat{\beta}$ is not unique but 
this is the simplest choice which is self adjoint (others have been considered in the literature, e.g. \cite{ms}).  For the rest of this paper we
use this as our approximate $\beta$ operator and drop the subscript $\Delta$. Its 
action on our basis kets is given by,
\begin{eqnarray}
	\label{eq:26}
	\hat{\beta}|v;p_\phi\rangle &=& \frac{1}{i\sqrt{\Delta}}\left(|v+1;p_\phi\rangle - |v-1;p_\phi\rangle \right) .
\end{eqnarray}
The action of the other operators is straightforward, $\hat{V}$ and $\hat{p}_\phi$ act by
multiplication,
\begin{eqnarray}
	\label{eq:27}
	\hat{V}|v;p_\phi\rangle &=& \left(\frac{8\pi\gamma}{6}\right)^\frac{3}{2}\frac{l_p^3}{K}v|v;p_\phi\rangle , \\
	\label{eq:28}
	\hat{p}_\phi|v;p_\phi\rangle &=& p_\phi|v;p_\phi\rangle ,
\end{eqnarray}
and since we work in the $p_\phi$ representation $\phi$ acts by differentiation
\begin{equation}
	\label{eq:29}
	\hat{\phi} = -\frac{\hbar}{i}\frac{\partial}{\partial p_\phi} .
\end{equation}
\subsection{Restrictions on Parameters in Coherent State}\label{sec:R}
Before we move to computation we impose some physically motivated restrictions
on the parameters appearing in the coherent state \eqref{eq:23}. The first pair
of restrictions,
\begin{eqnarray}
	\label{eq:30}
	v' \gg 1	, \\
	\label{eq:31}
	\sqrt{\Delta}\beta' \ll 1 ,
\end{eqnarray} 
corresponds to late times $V' \gg l_p^3$ and $\dot{a} \ll 1$ respectively.  Namely,
that the scale factor be much larger than the Planck length and that the rate of change of
the scale factor be much smaller than the speed of light.  We will see later that \eqref{eq:31}
also holds well even at early times.

The next pair of restrictions demands that the spreads in $\hat{V}$ and $\hat{\beta}$ be small,
$\frac{\Delta V}{V} \ll 1$ and $\frac{\Delta \beta}{\beta} \ll 1$, or equivalently:
\begin{eqnarray}
	\label{eq:32}
	v'\epsilon \gg 1, \\
	\label{eq:33}
	\epsilon \ll \sqrt{\Delta}\beta'.
\end{eqnarray}
The last pair of restrictions on parameters demands that the spreads in $\phi$ and $p_\phi$ are small,
$\frac{\Delta \phi}{\phi} \ll 1$ and $\frac{\Delta p_\phi}{p_\phi} \ll 1$, or equivalently:
\begin{eqnarray}
	\label{eq:34}
	\phi \gg \epsilon_\phi, \\
	\label{eq:35}
	p_\phi \epsilon_\phi \gg 1.
\end{eqnarray}
We use these physically motivated restrictions in our calculations in the remainder of this 
paper. We now return to showing that $\psi$ is sharply peaked.  
\subsection{Verifying that $\psi$ is Sharply Peaked}\label{sec:VPSP}
Now that we have a candidate semiclassical state, we can calculate the expectation values
of the canonical variables and verify that the state $\psi$ is sharply peaked at classical
values of the canonical variables.

Calculating $\langle\hat{\beta}\rangle$ we obtain,
\begin{equation}
	\label{eq:36}
	\langle\hat{\beta}\rangle = \frac{2}{\sqrt{\Delta}}e^{-\frac{1}{4}\epsilon^2}\sin(\half\sqrt{\Delta}\beta').
\end{equation}
Thus we must take $\epsilon \ll 1$ for $\langle\hat{\beta}\rangle$ to approximately 
agree with the classical $\beta$. Similarly,
\begin{equation}
	\label{eq:37}
	\langle\hat{\beta}^2\rangle = \frac{2}{\Delta}\left[1-e^{-\epsilon^2}\cos(\sqrt{\Delta}\beta')\right] .
\end{equation}
Thus,
\begin{equation}
	\label{eq:38}
	\Delta \beta^2 \simeq \frac{2\epsilon^2}{\Delta} ,
\end{equation}
where to get the last line we have used $\cos(\sqrt{\Delta}\beta') \simeq 1$ 
and $\epsilon \ll 1$ .  Now we obtain the expectation value and spread of $V$,
\begin{equation}
	\label{eq:39}
	\langle\hat{V}\rangle \simeq V' ,
\end{equation}
where to obtain the last line we have done Poisson resummation on the sum over $n$.  Similarly,
\begin{eqnarray}
	\label{eq:40}
	\langle \hat{V}^2\rangle &=& V'^2 + \frac{1}{2\epsilon^2}\frac{l_p^6}{K^2}\left(\frac{8\pi\gamma}{6}\right)^3 , \\
	\label{eq:41}
	\Delta V^2 &=& \frac{1}{2\epsilon^2}\frac{l_p^6}{K^2}\left(\frac{8\pi\gamma}{6}\right)^3 .
\end{eqnarray}
Thus, up to the approximation used to arrive at \eqref{eq:38} the product of uncertainties
is the minimum possible,
\begin{equation}
	\label{eq:42}
	\Delta \beta \Delta V \simeq \frac{\kappa\gamma}{2}\frac{\hbar}{2}.
\end{equation}

We make a brief detour to show that we can satisfy both of the conditions \eqref{eq:32} 
and \eqref{eq:33} on the phase space if the width $\epsilon$ of the coherent state is
properly chosen as a function of the phase space point. Notice that \eqref{eq:32} requires that
$\epsilon$ satisfy $v'\epsilon \gg 1$ and \eqref{eq:33} requires that $\epsilon \ll \sqrt{\Delta}\beta$,
but $\sqrt{\Delta}\beta \ll 1$ so it appears that there is a tension between these 2
conditions.  Nonetheless, if $p_\phi'$ is large, which is reasonable for a universe that 
would increase to macroscopic size then we can choose $\epsilon(v')$ such that both of the 
conditions $\frac{\Delta V}{V}$ and $\frac{\Delta\beta}{\beta}$ are satisfied. Looking 
at the relative uncertainties, we get
\begin{eqnarray}
	\frac{\Delta V}{V} &=& \frac{1}{\sqrt{2}\epsilon V'}\frac{l_p^3}{K}\left(\frac{8\pi\gamma}{6}\right)^3 \ll 1\\
	\frac{\Delta\beta}{\beta} &=& \frac{\sqrt{2}\epsilon}{\Delta\beta'} \ll 1
\end{eqnarray}
The first condition shows that we need to choose $\epsilon$ as a function of the phase phase
variable $V'$ s.t. $\epsilon=\frac{\lambda}{v'}$ (for $\lambda \ll (\frac{8\pi\gamma}{6})^3K^{-1}$)
in order to satisfy the inequality.  As for the 2nd condition we use the the classical form for 
$\beta^2=\frac{\kappa\gamma^2}{3}\rho$, where $\rho=\frac{p_\phi^2}{2V^2}$ for a scalar field, to 
show that the inequality is equivalent to taking $\sqrt{3\pi}\hbar \ll p_\phi$.  For a universe
that grows to macroscopic size such that $p_\phi$ satisfies this then we can choose $\lambda$ 
in such a way to satisfy $\lambda \ll (\frac{8\pi\gamma}{6})^3K^{-1}$, e.g. $\lambda=70, p_\phi=5000$.
Therefore we can satisfy these two conditions simultaneously.

Continuing, taking the expectation values of $\hat{\phi}$ and $\hat{p}_\phi$ yields, 
\begin{eqnarray}
	\label{eq:43}
	\langle \phi\rangle = \phi' ,
\end{eqnarray}
and 
\begin{eqnarray}
	\label{eq:44}
	\langle \phi^2\rangle = \phi'^2 + \frac{1}{2}\epsilon_\phi^2 ,
\end{eqnarray}
so
\begin{eqnarray}
	\label{eq:45}
	\Delta \phi = \frac{\epsilon_\phi}{\sqrt{2}} .
\end{eqnarray}
Thus, to have $\frac{\Delta \phi}{\phi} \ll 1$ we must have $\epsilon_\phi \ll \phi$.  Similarly for $p_\phi$ and $p_\phi^2$,
\begin{eqnarray}
	\label{eq:46}
	\langle  p_\phi\rangle = p_\phi', \\
	\label{eq:47}
	\langle  p_\phi^2\rangle = p_\phi'^2 + \frac{1}{2\epsilon_\phi^2},
\end{eqnarray}
therefore we have for $\Delta p_\phi^2$,
\begin{eqnarray}
	\label{eq:48}
	\Delta p_\phi = \frac{1}{\sqrt{2}\epsilon_\phi}.
\end{eqnarray}
Thus, to have $\frac{\Delta p_\phi}{p_\phi} \ll 1$ we must take $p_\phi \epsilon_\phi \gg 1$ .

Notice that this coherent state saturates the Heisenberg uncertainty bound as it should),
\begin{equation}
	\label{eq:49}
	\Delta \phi \Delta p_\phi = \frac{1}{2}.
\end{equation}
Therefore we have confirmed that the expectation values of the operators
corresponding to the canonical variables agree with their classical values.
Additionally, we have also shown that this coherent state saturates the Heisenberg
bound, as it should.  Therefore it is a natural kinematical semiclassical state to work 
with in order to obtain these `effective equations' from the geometric quantum mechanics
framework.

\subsection{Expectation Value of the Hamiltonian Constraint Operator}\label{EVHCO}
Having verified that $\psi$ is indeed sharply peaked around classical values
of the canonical variables, we proceed to calculate the expectation value of the
Hamiltonian constraint operator.  Recall that, if we can find an approximately horizontal
section of the Hilbert space then the effective quantum dynamics on that section is
generated by the effective Hamiltonian which is just the expectation value of the 
Hamiltonian constraint operator.  We calculate this expectation value now, leaving the
proof that the section is approximately horizontal to a later section.

Now notice that the self-adjoint constraint is given by
\begin{eqnarray}
	\label{eq:50}
	C &=& \frac{1}{16\pi G}C_{grav}+\half C_\phi
\end{eqnarray}
where
\begin{eqnarray}
	\label{eq:51a}
	\hat{C}_{grav} &=& \sin(\bar{\mu}c)\left\{\frac{24i\sgn(\mu)}{8\pi\gamma^3\bar{\mu}^3l_p^2}\left[\sin\left(\frac{\bar{\mu}c}{2}\right)\hat{V}\cos\left(\frac{\bar{\mu}c}{2}\right)\right.\right.\nonumber \\
	&&\left.\left.-\cos\left(\frac{\bar{\mu}c}{2}\right)\hat{V}\sin\left(\frac{\bar{\mu}c}{2}\right)\right]\right\}\sin\left(\bar{\mu}c\right) \\
	\label{eq:51b}
	\hat{C}_\phi &=& \widehat{p^{-\frac{3}{2}}}\otimes p_\phi^2
\end{eqnarray}
The action of the these on our basis kets is given by,
\begin{eqnarray}
	\label{eq:52}
	\hat{C}_{grav}|v;p_\phi\rangle &=& f_+(v)|v+4;p_\phi\rangle + f_0(v)|v;p_\phi\rangle \nonumber \\
	&&+ f_-(v)|v-4;p_\phi\rangle \\
	\hat{C}_\phi|v;p_\phi\rangle &=& \half\left(\frac{6}{8\pi\gamma l_p^2}\right)^\frac{3}{2}B(v)p_\phi^2|v;p_\phi\rangle \\
\end{eqnarray}
where
\begin{eqnarray}
	\label{eq:53}
	f_+(v) &=& \frac{27}{16}\sqrt{\frac{8\pi}{6}}\frac{Kl_p}{\gamma^\frac{3}{2}}|v+2|||v+1|-|v+3|| \\
	f_-(v) &=& f_+(v-4) \\
	f_0(v) &=& -f_+(v)-f_-(v)
\end{eqnarray}
and
\begin{equation}
	\label{eq:54}
	B(v)=\left(\frac{3}{2}\right)^3K|v|||v+1|^\third-|v-1|^\third|^3
\end{equation}
and
\begin{equation}
	\label{eq:54b}
	\widehat{p^{-\frac{3}{2}}}|v;p_\phi\rangle = \left(\frac{6}{8\pi\gamma l_p^2}\right)^\frac{3}{2}B(v)|v;p_\phi\rangle
\end{equation}
Using these one can go ahead and compute the expectation value of the Hamiltonian constraint
$\langle\hat{C}\rangle$.  Note that such a summation includes a summation over negative 
$v$ even though we are considering late times $v\gg 1$.  However the contribution introduced 
from these terms is exponentially suppressed.  For a proof of this see \cite{jw} and the appendix in \cite{amj}.  
We can evaluate the sum by Poisson resummation and find an asymptotic expansion for the first term
in the series obtaining the effective constraint to leading and subleading order.  
\begin{widetext}
\begin{eqnarray}
	\label{eq:55}
	\langle \hat{C}\rangle &=& -\frac{3}{16\pi G\gamma^2\bar{\mu}'^2}p^\half\left[1+e^{-4\epsilon^2}\left(2\sin^2(\sqrt{\Delta}\beta')-1\right)\right] + \half\left(p_\phi'^2+\frac{1}{2\epsilon_\phi^2}\right)\left(\frac{6}{8\pi\gamma l_p^2}\right)^\frac{3}{2}K\left[\frac{1}{v'} + O(v'^{-3},v'^{-3}\epsilon^{-2})\right]\nonumber \\
\end{eqnarray}
\end{widetext}

\subsection{Equations of Motion}\label{EoM}
Recall that to evaluate $\dot{O}$ for some operator $O$ we simply take the commutator 
between $O$ and the Hamiltonian and divide by $i\hbar$.  We begin with $\langle\dot{\beta}\rangle$,
\begin{equation}
	\label{eq:61}
	\langle \dot{\beta} \rangle =\frac{\langle \psi|\frac{1}{i\hbar}\left(\frac{1}{16\pi G}[\beta,C_{grav}]+\half[\beta,C_\phi]\right)|\psi\rangle}{\langle\psi|\psi\rangle} .
\end{equation}	
which can be computed using a similar expansion to that used to find the
expectation value of the Hamiltonian constraint to yield,
\begin{widetext}
\begin{eqnarray}
	\label{eq:63}
	\langle \dot{\beta} \rangle &\simeq & -\frac{1}{16\pi G}\frac{27}{16\hbar}\left(\frac{8\pi \gamma}{6}\right)^\half\frac{Kl_p}{\gamma^2\sqrt{\Delta}}\left[4e^{-\frac{25}{4}\epsilon^2}\cos\left(\frac{5}{2}\sqrt{\Delta}\beta'\right)+4e^{-\frac{9}{4}\epsilon^2}\cos\left(\frac{3}{2}\sqrt{\Delta}\beta'\right) - 8e^{-\frac{1}{4}\epsilon^2}\cos\left(\half\sqrt{\Delta}\beta'\right)\right] \nonumber \\
	&&- \left(p_\phi'^2+\frac{1}{2\epsilon_\phi^2}\right)\frac{1}{2V'^2}
\end{eqnarray}
\end{widetext}
Now for $\langle \dot{V} \rangle$
\begin{equation}
	\label{eq:64}
	\langle \dot{V} \rangle =\frac{\langle \psi|\frac{1}{i\hbar}\left(\frac{1}{16\pi G}[V,C_{grav}]+\half[V,C_\phi]\right)|\psi\rangle}{\langle\psi|\psi\rangle}
\end{equation}	
Notice, $V$ commutes with $C_\phi$ so there is no contribution 
dependent on the scalar field and we just retain the contribution from the
commutator with $C_{grav}$ which yields
\begin{equation}
	\label{eq:65}
	\langle\dot{V}\rangle \simeq \frac{3V'}{\gamma}e^{-4\epsilon^2}\frac{\sin(2\sqrt{\Delta}\beta')}{2\sqrt{\Delta}} 
\end{equation}
Now focus on $\langle\dot{\phi}\rangle$ and $\langle\dot{p}_\phi\rangle$,
\begin{eqnarray}
	\label{eq:66}
	\langle \dot{\phi} \rangle = \frac{\langle \psi|\frac{1}{2i\hbar}[\phi,C_\phi]|\psi\rangle}{\langle\psi|\psi\rangle} ,
\end{eqnarray}
since $\phi$ commutes with $C_{grav}$
\begin{eqnarray}
	\label{eq:67}
	\langle \dot{\phi} \rangle &\simeq & \frac{p_\phi'}{V'}+O\left(\frac{1}{V'^3}\right)
\end{eqnarray}
Similarly,
\begin{eqnarray}
	\label{eq:68}
	\langle \dot{p_\phi} \rangle &=& \frac{\langle \psi|\frac{1}{i\hbar}\left(\frac{1}{16\pi G}[p_\phi,C_{grav}]+\half[p_\phi,C_\phi]\right)|\psi\rangle}{\langle\psi|\psi\rangle} = 0 \nonumber \\
\end{eqnarray}
since $p_\phi$ commutes with both $C_{grav}$ and $C_\phi$.

From these expressions we obtain the 'effective equations'.
\section{Effective Equations}\label{sec:EE}
Recall that in the geometric quantization picture we take as our basic observables, the expectation values.  
\begin{eqnarray}
	\label{eq:74}
	\bar{\beta} &=& \langle\beta\rangle = \frac{2}{\sqrt{\Delta}}e^{-\frac{1}{4}\epsilon^2}\sin(\half\sqrt{\Delta}\beta') \\
	\label{eq:75}
	\bar{V} &=& \langle V\rangle = V' \\ 
	\label{eq:76}
	\bar{\phi} &=& \langle \phi\rangle = \phi' \\
	\label{eq:77}
	\bar{p}_\phi &=& \langle p_\phi\rangle = p_\phi'
\end{eqnarray}
Recall, that the coordinates on the classical phase space are the expectation values, thus
these barred variables are the coordinates on the classical phase space.  So we search
for an effective description in terms of these variables.

We thus invert these \eqref{eq:74} - \eqref{eq:77} for the primed variables and express 
the evolution equations in terms of the barred variables.  We look at the first few terms 
of these asymptotic expansions to get the leading and next to leading behavior and apply 
the approximations listed above in \ref{sec:R}.  Doing this we obtain the 'effective 
equations of motion'  (the main result of this paper),
\begin{eqnarray}
	\label{eq:78}
	\bar{C} &=& -\frac{3}{\kappa\gamma^2}\bar{V}\bar{\beta}^2\left(1-\frac{1}{4}\Delta\bar{\beta}^2\right)-\frac{6\epsilon^2}{\kappa\gamma^2}\frac{\bar{V}}{\Delta}\nonumber \\
	&&+\frac{\bar{p}_\phi^2}{2\bar{V}}\left[1+O(\bar{V}^{-2},\bar{V}^{-2}\epsilon^{-2})\right] \\
	\label{eq:79}
	\dot{\bar{\beta}} &=& \frac{3}{4\gamma}\sqrt{1-\frac{1}{4}\Delta\bar{\beta}^2}\left[-2\bar{\beta}^2+\Delta\bar{\beta}^4\right]\nonumber \\
	&&-\frac{\kappa}{4}\sqrt{1-\frac{1}{4}\Delta\bar{\beta}^2}\frac{\bar{p}_\phi^2}{V'^2}[1+O(\bar{V}^{-2},\bar{V}^{-2}\epsilon^{-2})]\\
	\label{eq:80}
	\dot{\bar{V}} &=& 3\frac{\bar{\beta}}{\gamma}\bar{V}\sqrt{1-\frac{\Delta\bar{\beta}^2}{4}}\left(1-\half\Delta\bar{\beta}^2\right)\\
	\label{eq:81}
	\dot{\bar{\phi}} &=&  \frac{\bar{p}_\phi}{\bar{V}}+O(\bar{V}^{-3})\\
	\label{eq:82}
	\dot{\bar{p}}_\phi &=& 0
\end{eqnarray}

As discussed in the Appendix, Eqs (\eqref{eq:78} and \eqref{eq:80} imply that the Hubble parameter $H:= \frac{\dot{a}}{a} \equiv \frac{\dot{V}}{V}$
satisfies an effective Friedmann equation,
\begin{equation}
	H^2=\frac{\kappa\rho}{3}\left(1-\frac{\rho}{\rho_{crit}}\right)+O(\epsilon^2)
\end{equation}
where 
\begin{equation}
	\rho_{crit}=\frac{3}{\kappa\gamma^2\Delta}
\end{equation}
which incorporates the leading quantum corrections. Somewhat surprisingly, as numerical simulations show, these corrections are already 
sufficient to correctly reproduce the main features of full quantum dynamics. These equations have been used in the literature to make 
certain phenomenological predictions (see,e.g., \cite{mss}). 

One may ask, what is the error involved in making the approximations that led to equations
\eqref{eq:78}-\eqref{eq:82}.  In these equations we are keeping corrections of
$O(\Delta\beta^2)$ relative to the classical expressions but ignoring terms of $O(\Delta^2\beta^4)$.
How much of an error is one making by ignoring these terms?  Let us compute, $\Delta\beta^2$,
\begin{equation}
	\label{eq:86}
	\Delta\beta^2=\Delta\gamma^2\frac{\dot{a}^2}{a^2}=\frac{\rho}{\rho_c}\left(1-\frac{\rho}{\rho_c}\right)
\end{equation}
Notice, this function starts off at 0 at late times since $\rho/\rho_c \sim 0$, it slowly increases
reaches a maximum value of $1/4$ at $\rho=\rho_c/2$ and then again goes to $0$ at the point 
$\rho/\rho_c=1$.  Therefore, at the worst we are making an error of about $6\%$ in
ignoring terms of the order $\Delta^2\beta^4$ and this occurs at $\rho=\rho_c/2$.  At other times,
i.e. at late times and at times near the bounce point, this approximation is very good.

Additionally, one can pullback the symplectic structure $\Omega$ to our candidate horizontal section
and analyze the dynamics generated by this effective Hamiltonian.  Does it give the equations 
\eqref{eq:79}-\eqref{eq:82}? One can pullback $\Omega$ and verify that it is given by
\begin{equation}
	\label{eq:ss}
	\Omega = \frac{2}{\kappa\gamma}\frac{1}{\sqrt{1-\frac{1}{4}\Delta\bar{\beta}^2}}d\bar{\beta}\wedge d\bar{V}+d\bar{\phi} \wedge d\bar{p}_\phi 
\end{equation}
and that its associated Poisson bracket is
\begin{eqnarray}
	\label{eq:pb}
	\{f,g\} &=& \frac{\kappa\gamma}{2}\sqrt{1-\frac{1}{4}\Delta\bar{\beta}^2}\left(\frac{\partial f}{\partial\bar{\beta}}\frac{\partial g}{\partial \bar{V}}-\frac{\partial g}{\partial\bar{\beta}}\frac{\partial f}{\partial\bar{V}}\right) \nonumber \\
		    &+& \frac{\partial f}{\partial \bar{\phi}}\frac{\partial g}{\partial \bar{p}_\phi}-\frac{\partial g}{\partial\bar{p}_\phi}\frac{\partial f}{\partial \bar{\phi}}
\end{eqnarray}

We can verify that our `effective equations' are consistent by checking that the
Poisson brackets in terms of the barred variables hold, to within our order of approximation.
That is, $(\bar{\beta},\bar{V},\bar{\phi},\bar{p}_\phi)$ are the coordinates on the horizontal
section, $\Gamma$, and we wish to know whether they are preserved, i.e. is the vector 
$(\dot{\bar{\beta}},\dot{\bar{V}},\dot{\bar{\phi}},\dot{\bar{p}}_\phi)$ tangent to $\Gamma$
or off it?  If it is not tangent to $\Gamma$ then is it approximately tangent?  That is,
are its components off $\Gamma$ small relative to the tangential components?  This is indeed 
the case, since to our order of approximation the equations of motion in terms of the barred 
variable hold.  Indeed, it can be verified that the following equations hold in terms of the 
pullback of the symplectic structure to the barred variables.
\begin{eqnarray}
	\label{eq:83}
	\dot{\bar{\beta}} &=& \left\{\bar{\beta},\bar{C}\right\} \nonumber \\
	\dot{\bar{V}} &=& \left\{\bar{V},\bar{C}\right\} \nonumber \\
	\dot{\bar{\phi}} &=& \left\{\bar{\phi},\bar{C}\right\} \nonumber \\
	\dot{\bar{p}}_\phi &=& \left\{\bar{p}_\phi,\bar{C}\right\} 
\end{eqnarray}
This is a highly nontrivial consistency check on the formalism as well as approximations used
at intermediate steps.

\section{Conclusion}\label{sec:C}
Thus, we have constructed kinematical quantum states that closely approximate
solutions to the classical Einstein equations.  Therefore, at least for the case
of a Friedmann universe with a free scalar field in the context of loop quantum
cosmology, it is the case that there exist suitable quantum states that closely
approximate solutions to the classical Einstein equations and \textit{remain}
sharply peaked along a quantum corrected, semiclassical trajectory.

Furthermore we note, that we have shown that the classical Einstein's equations are not just
reproduced identically in the quantum theory but they do indeed pick up corrections due 
to quantum effects.  In this work, we have used the geometric quantum mechanics
framework and approximated the full quantum dynamics in the (infinite dimensional) 
Hilbert space by a system of `effective equations', incorporating the leading quantum 
corrections, on a finite dimensional submanifold isomorphic to the classical phase space. 

\section{Acknowledgments}\label{sec:Ack}
The author would like to thank Abhay Ashtekar for his continuing support, originally 
suggesting this project and for discussions; Martin Bojowald for reading a draft of this paper and for 
discussions; and Parampreet Singh and Kevin Vandersloot for discussions.  While 
working on this the author was supported in part by the Bunton-Waller Assistantship funds, 
 funds at Penn State, and by the Alfred P. Sloan foundation.

\section{Appendix: Validity of Effective Equations Near the Bounce}\label{sec:App}
It is instructive to write the `effective equations' in standard form.  Therefore,
in this appendix we address the question: what is the form of the corrected Friedmann and Raychaudhuri equations?

We can obtain the corrected Friedmann equation by, working to next-to leading order,
solving \eqref{eq:78} for $\bar{\beta}$, 
substituting into \eqref{eq:80}, and then computing the Hubble parameter $H=\frac{\dot{a}}{a}=\frac{\dot{V}}{3V}$,
\begin{equation}
	\label{eq:84}
	H^2=\frac{\kappa\rho}{3}\left(1-\frac{\rho}{\rho_{crit}}\right)+O(\epsilon^2)
\end{equation}
where 
\begin{equation}
	\label{eq:85}
	\rho_{crit}=\frac{3}{\kappa\gamma^2\Delta}
\end{equation}
Notice, in addition to recovering the classical $\frac{\kappa}{3}\rho$ term we get a
correction term that becomes important when the density $\rho$ becomes very large.
We see that the minus sign allows $H$, and hence $\dot{a}$ to be zero.  Thus allowing the 
possibility of a bounce.  Thus, even though these equations hold at late times since
we are working in the semiclassical regime, we already see that gravity is becoming
repulsive and allowing the scale factor to bounce when we evolve backwards in time
and approach the classical big bang singularity.  

On \eqref{eq:84} we must make two comments.  First, a priori one would not expect 
\eqref{eq:84} to describe the correct dynamics near the bounce point because the bounce
point $\rho/\rho_c =1$ lies outside of the regime of our approximation. That is, the
natural domain of applicability of these `effective equations' is a late-time, large-volume 
approximation and the point $\rho/\rho_c=1$ lies well outside of that regime since the 
condition \eqref{eq:33} fails badly at that point. However, numerical work (see reference 
2 in \cite{aps}) has shown that the dynamics is indeed described well by \eqref{eq:84} even at the bounce and hence this approximation and the 
results obtained in this work continue to be good even beyond their expected regime.  Secondly,
near the bounce point the term in parentheses in \eqref{eq:84}is approaching 0, but since the approximation \eqref{eq:33}
is breaking down there it is not clear that the $O(\epsilon^2)$ term is negligible there.  Thus
\eqref{eq:84} appears to break down there, but the numerical work has shown that it holds very well
with negligible $O(\epsilon^2)$ corrections.  As is not uncommon in physics, the effective theory
is valid well beyond the domain for which it was constructed.

So now we have the effective Friedmann equation in this model.  It would be instructive, for completeness,
to also obtain the conservation equation and the corrected Raychaudhuri equation for this effective
theory.  Recalling that $\rho=p$ for a massless scalar field, classically these equations are
\begin{eqnarray}
	\label{eq:83b}
	\dot\rho &=& -6\frac{\dot{a}}{a}\rho, \\
	\label{eq:83c}
	3\frac{\ddot{a}}{a} &=& -2\kappa\rho
\end{eqnarray}
Let us compute the corrected versions of these equations.  Using the modified
Poisson bracket, we can compute $\dot{\rho}=\{\rho,\bar{C}\}$.  Doing this we obtain
precisely \eqref{eq:83b}, so the continuity equation is not modified.  Similarly,
we can take \eqref{eq:80} and compute $\ddot{\bar{V}}=\{\dot{\bar{V}},\bar{C}\}$,
expressing it in terms of the scale factor $a$, and solving for $3\frac{\ddot{a}}{a}$.
One obtains
\begin{equation}
	\label{eq:84d}
	3\frac{\ddot{a}}{a}=-2\kappa\rho\left(1-\frac{5}{2}\frac{\rho}{\rho_c}\right)+O(\epsilon^2)
\end{equation}
Notice that classically this expression is always negative.  However, now there is a correction of $O(\frac{\rho}{\rho_c})$. 
Notice, that the corrected Friedmann equation \eqref{eq:84} tells us there is 
a bounce at $\rho/\rho_c=1$.  At that density the corrected Raychaudhuri equation
\eqref{eq:84d} tells us that $\ddot{a}$ is positive as it should be if there is to be 
a bounce.

\end{document}